\documentclass[aps,pra,twocolumn,superscriptaddress]{revtex4-2}
\usepackage{amsmath,amssymb,amsfonts}
\usepackage{graphicx}
\usepackage{bm}
\usepackage{microtype}
\usepackage{hyperref}

\begin{document}

\title{High-fidelity multiqubit gates with Rydberg atoms via level-crossing-free Rapid adiabatic passage}

\author{Yichi Zhang}
\email{zhangyichi@sxu.edu.cn}
\author{Ximo Wang}
\author{Zhenqi Bai}
\author{Xu Zhao}
\author{Tiecheng Wang}
\affiliation{College of Physics and Electronic Engineering, Shanxi University, 030006 Taiyuan, People’s Republic of China}
\affiliation{Collaborative Innovation Center of Extreme Optics, Shanxi University, Taiyuan, Shanxi 030006, People’s Republic of China}

\date{\today}

\begin{abstract}
We propose a rapid adiabatic passage scheme based on level-crossing-free pulses for generating multiqubit entangled states in Rydberg-atom systems. Unlike conventional rapid adiabatic passage (RAP) schemes that rely on level crossings, our approach employs an antisymmetric Rabi frequency and an even-symmetric detuning, enabling robust population transfer without passing through any level crossing. We exploit the Rydberg blockade effect to deterministically prepare entangled states from an initial product state. Specifically, two sequential RAP pulses separated by a $\pi_g$ pulse generate two-qubit Bell states, three-qubit W states, four-qubit GHZ states, and six-qubit honeycomb W states. Our numerical simulations show that the fidelities exceed $0.9997$ for the two-qubit Bell state and the three-qubit W state, reach $0.997$ for the four-qubit GHZ state, and surpass $0.9995$ for the six-qubit honeycomb W state. The scheme demonstrates excellent robustness against pulse parameter fluctuations, with the fidelity remaining above $0.99$ within $\pm 5\%$ parameter variation. This scheme provides a simple, efficient, and robust method for entangled state preparation in neutral-atom quantum information processing.
\end{abstract}

\maketitle

\section{Introduction}

Quantum entanglement is one of the most distinctive phenomena in quantum mechanics and a core resource for quantum information science. In quantum computing, entangled states are essential for realizing quantum parallel computation and quantum error correction \cite{evered2023, bluvstein2024, ma2023, scholl2023, wu2022, henriet2020, singh2023}; in quantum communication, they underpin quantum key distribution, quantum teleportation, and quantum networks \cite{huie2021, wehner2018}; in quantum metrology, entangled states can surpass the standard quantum limit and achieve measurement precision beyond classical bounds \cite{degen2017, marciniak2022}. Efficient and high-fidelity preparation and manipulation of quantum entangled states is therefore a key prerequisite for advancing quantum information technology from the laboratory to practical applications.

Among various physical platforms, neutral-atom systems based on Rydberg atoms have attracted considerable attention due to their unique advantages \cite{Bharti2023, saffman2010, barredo2016, shi2022, browaeys2020, morgado2021, wu2021, kaufman2021, wintersperger2023}. First, Rydberg atoms possess strong long-range interactions, with interaction strengths between two atoms separated by several micrometers reaching hundreds of MHz, far exceeding typical laser Rabi frequencies (about $0.1$--$1$ MHz). Second, Rydberg states have relatively long lifetimes (about $10$--$100~\mu$s), providing sufficient time windows for quantum manipulation. Most importantly, the Rydberg blockade effect---within the blockade radius, only one atom can be excited to the Rydberg state---provides a natural physical mechanism for realizing deterministic quantum logic gates and entangled state preparation \cite{Liang2025, Wang2026, urban2009, wilk2010, isenhower2010, madjarov2020, jo2020, finkelstein2024, picken2018}. In recent years, optical tweezer technology has enabled precise manipulation and arbitrary arrangement of individual atoms \cite{barredo2016, madjarov2019, ebadi2021, scholl2021, pause2024, manetsch2025, tao2024}, laying the foundation for constructing large-scale Rydberg atom quantum processors.

Rapid adiabatic passage (RAP) is a quantum control technique that utilizes time-varying laser fields to achieve complete population transfer \cite{jaksch2000, muller2011, shore2017, guerin2003}. Its core idea is to slowly change the laser frequency, causing the system to undergo a level crossing. Under adiabatic conditions, atoms follow the instantaneous eigenstates and thereby achieve complete transfer from the initial state to the target state. RAP technology possesses natural robustness against pulse parameter fluctuations, which makes it valuable in quantum information processing. In recent years, RAP has been widely applied in Rydberg atom systems for realizing quantum logic gates and entangled state preparation \cite{beterov2020, mitra2020, saffman2020, sun2020, pelegri2022, fu2022, chang2023}. However, conventional RAP schemes rely on precise level crossings, and at the level crossing point the instantaneous eigenenergy gap of the system is extremely small, making nonadiabatic transitions highly probable and thereby limiting the theoretical ceiling of manipulation fidelity. In many-body systems, complex energy level structures may lead to multiple level crossings interfering with each other, introducing nonadiabatic transitions and reducing manipulation fidelity. Moreover, conventional RAP schemes have relatively strict requirements on pulse shape, timing, and parameters \cite{shi2022, yang2019, robicheaux2021, pagano2022, fromonteil2023, mohan2023, jandura2023, xiong2026}, and different qubit-number systems require separate parameter optimization, which increases the experimental complexity.
\begin{figure}[htbp]
	\centering
	\includegraphics[width=0.5\textwidth]{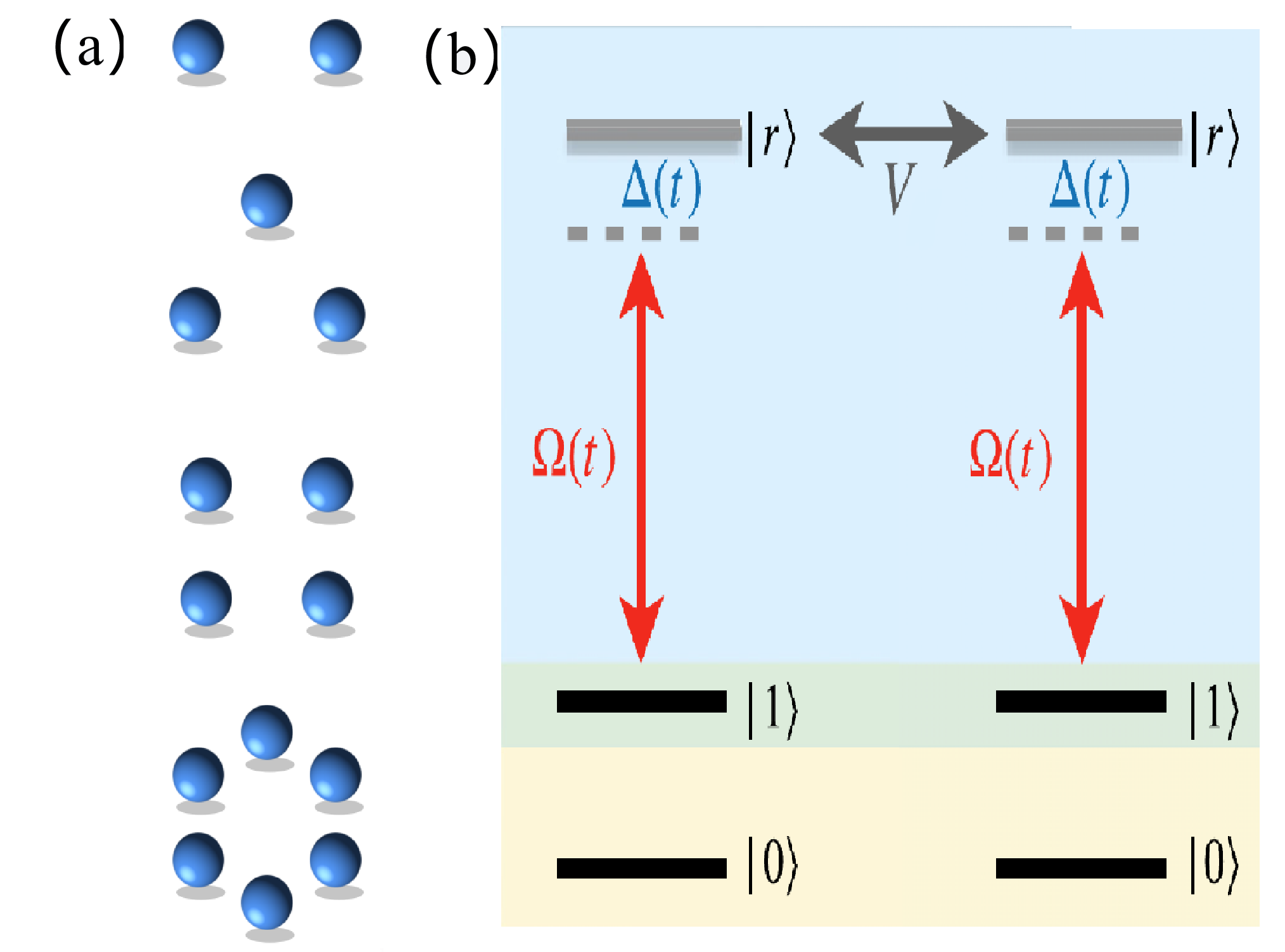}
	\caption{Schematic of the theoretical model. (a) Schematic of the possible atomic arrangements considered in this work. (b) Atomic-level structure of two-atom qubits featuring Rydberg–Rydberg interactions between the $|r\rangle$ states.}
	\label{fig:1}
\end{figure}

In this paper, we employ level-crossing-free rapid adiabatic passage in Rydberg atoms to realize multiqubit entangled states. Our focus is on the preparation of two-qubit Bell states, three-qubit W states, four-qubit GHZ states, and six-qubit honeycomb W states. The level-crossing-free pulses adopt antisymmetric Rabi frequencies and even-symmetric detunings, ensuring that the system does not pass through any level crossing throughout the evolution. This design significantly reduces the risk of nonadiabatic transitions by keeping the detuning strictly positive and only touching resonance at the pulse center, thereby avoiding the dangerous region of minimal energy gap where the adiabatic condition is most vulnerable to violation. Compared with conventional RAP schemes, the proposed scheme exhibits significant advantages in fidelity, robustness, pulse simplicity, and parameter optimization complexity. The pulse form is simple, and there is no strong coupling between the parameters. Furthermore, the parameter optimization results can be directly applied to systems with different numbers of qubits, without requiring separate re-optimization.

The remainder of this paper is structured as follows. In Section~\ref{sec:theory}, we introduce the atomic model and the level-crossing-free pulse architecture. In Section~\ref{sec:numerical}, we present numerical results for the two-qubit Bell state, three-qubit W state, four-qubit GHZ state, and six-qubit honeycomb W state, and analyze the robustness under parameter fluctuations. Finally, conclusions are given in Section~\ref{sec:generalization and conclusion}.

\section{Theoretical Model and Scheme}\label{sec:theory}

\subsection{Atomic model}

We consider a system composed of $N$ identical atoms, as shown in Fig.~\ref{fig:1}(b), each having two hyperfine ground states $|0\rangle$ and $|1\rangle$ and a Rydberg excited state $|r\rangle$. Quantum information is encoded in the ground states $|0\rangle$ and $|1\rangle$. Through a global laser field, $|1\rangle$ and $|r\rangle$ are coupled with Rabi frequency $\Omega(t)$ and detuning $\Delta(t)$. Both atoms are subject to identical laser driving, and the system Hamiltonian can be written as (taking $\hbar = 1$)
\begin{equation}
\hat{H}(t) = \sum_{i=1}^{N} \left[\frac{\Omega(t)}{2}\left(|r\rangle_i\langle 1|_i + \text{h.c.}\right) + \Delta(t)|r\rangle_i\langle r|_i\right] + \hat{V},\label{eq:H}
\end{equation}
where $\hat{V} = \sum_{i<j} V_{ij} |r\rangle_i\langle r|_i \otimes |r\rangle_j\langle r|_j$ is the Rydberg interaction term, with $V_{ij} = C_6/r_{ij}^6$ being the van der Waals interaction strength between atoms $i$ and $j$, $C_6$ the van der Waals coefficient, and $r_{ij}$ the interatomic distance \cite{singer2005}. When $V_{ij}$ is much larger than the Rabi frequency, the doubly excited state $|rr\rangle$ is effectively blocked, and the system evolution is restricted to the single-excitation subspace.

We evaluate the state preparation performance using the fidelity defined as
\begin{equation}
F = |\langle\psi_{\text{tar}}|\hat{\rho}_f|\psi_{\text{tar}}\rangle|^2,
\end{equation}
where $|\psi_{\text{tar}}\rangle$ denotes the target entangled state of the multiqubit system, and $\hat{\rho}_f$ represents the final density operator obtained by evolving the initial state under the Hamiltonian $\hat{H}(t)$ given in Eq.~\eqref{eq:H} up to the total evolution time $\tau_{\text{tot}}$. To assess the robustness of the scheme under realistic experimental conditions, we also investigate dissipative effects through the Lindblad master equation \cite{manzano2020},
\begin{equation}
\frac{\partial\hat{\rho}(t)}{\partial t} = -i[\hat{H}(t), \hat{\rho}(t)] + \mathcal{L}[\hat{\rho}(t)],
\end{equation}
where $\mathcal{L}[\hat{\rho}] = \sum_k \mathcal{D}_k[\hat{\rho}]$ represents the Lindblad superoperator accounting for the coupling between the system and its environment, and the dissipators
\begin{equation}
\mathcal{D}_k[\hat{\rho}] = \hat{L}_k \hat{\rho} \hat{L}_k^\dagger - \frac{1}{2}\left(\hat{L}_k^\dagger \hat{L}_k \hat{\rho} + \hat{\rho} \hat{L}_k^\dagger \hat{L}_k\right)
\end{equation}
capture the incoherent processes such as spontaneous emission and dephasing of the Rydberg state. For $^{133}\mathrm{Cs}$ atoms, we employ the following operators: $\hat{L}_1 = \sqrt{\gamma_r/16}\,|0\rangle\langle r|$ and $\hat{L}_2 = \sqrt{\gamma_r/16}\,|1\rangle\langle r|$ to model spontaneous decay into the ground-state manifold, and $\hat{L}_3 = \sqrt{7\gamma_r/8}\,|r\rangle\langle r|$ to account for dephasing. The total decay rate is set to $\gamma_{r}=1/(540~\mu\mathrm{s})$~\cite{sibalic2017}.
\begin{figure*}[htbp]
	\centering
	\includegraphics[width=0.6\textwidth]{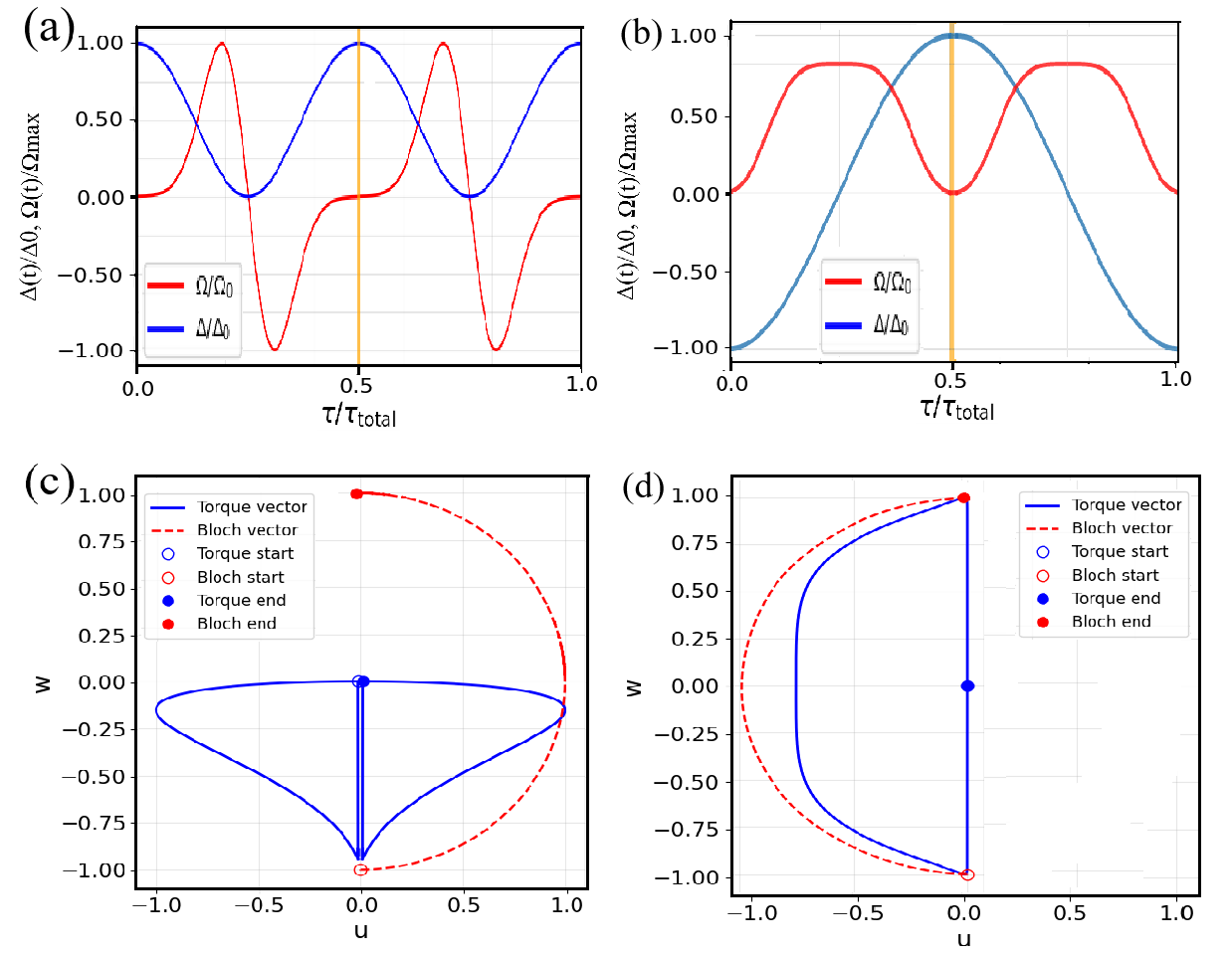}
	\caption{Schematic comparison of the level-crossing-free and conventional RAP pulses. (a) and (b)~Schematic illustrations of the level-crossing-free and level-crossing pulse schemes. (c) and (d)~Dynamics of the Bloch vector and the angular velocity vector projected onto the $(u,w)$ plane.
		Solid blue curves indicate the angular velocity vectors, and dashed red curves indicate the Bloch vectors, corresponding to panels~(a)~and~(b).}
	\label{fig:2}
\end{figure*}
\subsection{Level-crossing-free pulses}

We employ the pulse sequence illustrated in Fig.~\ref{fig:2}(a), where a $\pi_g$ pulse is incorporated into a continuous non-level-crossing pulse to generate entangled states. The time required for the $\pi_g$ pulse between the two pulses is negligible. Introducing the dimensionless time $\tau = \Omega_0 t$ (with $\Omega_0$ as the maximum Rabi frequency), the Rabi frequency and detuning during the $k$-th ($k = 1, 2$) pulse interval are taken as \cite{rangelov2010}
\begin{subequations}\label{eq:pulses}
\begin{align}
\Omega_k(\tau) &= -\Omega_0  C_\Omega\frac{\tau - \tau_k}{\tau_R}  \exp\left[-\left(\frac{\tau - \tau_k}{\tau_R}\right)^2\right], \label{eq:Omega}\\
\Delta_k(\tau) &= \Delta_0  C_\Delta\left(\frac{\tau - \tau_k}{\tau_D}\right)^2\exp\left[-\left(\frac{\tau - \tau_k}{\tau_D}\right)^2\right],\label{eq:Delta}
\end{align}
\end{subequations}
where $\tau_k = (2k-1)\tau_{\text{total}}/4$ denotes the pulse center, $\tau_R$ and $\tau_D$ control the temporal widths of the Rabi frequency and detuning, respectively, $\Delta_0$ is the detuning amplitude, and the normalization constants $C_\Omega = \sqrt{2e}$ and $C_\Delta = e$ ensure that the peak values reach $\Omega_0$ and $\Delta_0$, respectively.

The Rabi frequency in Eq.~\eqref{eq:Omega} is designed as a time-odd function satisfying the zero-area condition, which provides first-order suppression of laser intensity noise \cite{jiang2023, gungordu2022, vasilev2006}. Its time integral over a single pulse vanishes, forming a zero-area pulse\cite{lehto2016}. The detuning in Eq.~\eqref{eq:Delta} is a time-even function, satisfying $\Delta_k(\tau_k + t) = \Delta_k(\tau_k - t)$. It remains strictly positive throughout the evolution, keeping the system away from the resonance region without crossing zero. Our scheme employs two completely identical symmetric pulses. These features endow our scheme with distinctive physical attributes: the zero-area Rabi frequency pulse provides first-order suppression of laser intensity noise, significantly improving the robustness against experimental imperfections, while the strictly positive detuning ensures that the system always evolves with a large energy gap, effectively suppressing nonadiabatic transitions.

The specific pulse shape given by Eqs.~\eqref{eq:pulses} ensures that the system maintains smooth adiabatic evolution under different parameter combinations, without relying on a fixed resonance condition. This flexibility allows the same pulse architecture to be applied to systems with different numbers of qubits, simply by adjusting the pulse parameters, without requiring separate re-optimization of the pulse shape.

The pulse form of the conventional RAP is given in Ref.~\cite{xu2024},
\begin{subequations}\label{eq:6}
	\begin{align}
		\Omega(\tau) &= \frac{\Omega_{\max} \left( e^{-[(\tau - \tau_k)/\tau_R]^4} - a \right)}{1 - a}, \label{eq:6a}\\
		\Delta(\tau) &= (-1)^{k+1} \Delta_{\max} \sin\left[ \frac{\pi (\tau - \tau_k)}{\tau_D} \right],\label{eq:6b}
	\end{align}
\end{subequations}
where $(k-1)/2 \leqslant \tau/\tau_{\text{tot}} \leqslant k/2$, $\tau_k = (2k - 1)\tau_{\text{tot}}/4$, $k \in \{1, 2, \ldots\}$, and $a = \exp\left[-(\tau_{\text{tot}}/(4\tau_R))^4\right]$. The Rabi frequency $\Omega(t)$ in Eq.~\eqref{eq:6a} is designed as an even function of time that remains positive throughout, so that the time-integrated pulse area over a single pulse is nonzero. The detuning $\Delta(t)$ in Eq.~\eqref{eq:6b} adopts a sinusoidal chirp form, which varies continuously from negative to positive values within each chirp period and thus inevitably passes through the resonance point $\Delta = 0$. In the adiabatic limit, the system resides in a maximally coherent superposition state at this instant. Consequently, this conventional RAP pulse configuration is sensitive to fluctuations in the laser intensity.

Figure.~\ref{fig:2}(a) shows that the Rabi frequency $\Omega(\tau)$ of the zero-area pulse has an antisymmetric shape: in the first half-period, $\Omega(\tau)$ starts from zero, first increases positively and then decreases back to zero, in the second half-period, $\Omega(\tau)$ starts from zero, first increases negatively and then decreases back to zero. The detuning $\Delta(\tau)$ has an even-symmetric shape. This unique pulse shape ensures that the system achieves adiabatic evolution without passing through level crossings. The conventional RAP pulse scheme is shown in Fig.~\ref{fig:2}(b), where the detuning $\Delta(t)$ changes sign as a function of time. In conventional RAP schemes, the nonadiabatic coupling near level crossings is the main source of fidelity loss, whereas the zero-area scheme fundamentally eliminates this loss channel. Figures.~\ref{fig:2}(c) and~\ref{fig:2}(d) reveal two fundamentally distinct physical mechanisms for achieving complete population inversion via the level-crossing-free and conventional RAP schemes.

The level-crossing-free scheme proposed in this work demonstrates that efficient adiabatic transfer can be achieved by engineering a sign reversal of the Rabi frequency $\Omega(t)$ precisely at the instant when the detuning vanishes. Moreover, as shown in Fig.~\ref{fig:2}(a), the Rabi frequency $\Omega(t)$ employed here is a zero-area pulse whose time integral over a single pulse vanishes. This antisymmetric pulse shape provides first-order suppression of laser intensity noise, making the scheme inherently robust against fluctuations in the driving-field amplitude. 

The instantaneous eigenenergy gap is given by
\begin{align}
		\tilde{\Omega}(t) = \sqrt{\Omega^2(t) + \Delta^2(t)}
\end{align}
as shown in Fig.~\ref{fig:2}(c), since $\Delta(t)$ is even-symmetric and its square is always nonnegative, the adiabatic curves only touch but do not cross when $\Omega$ and $\Delta$ simultaneously approach zero. At this instant, the system resides exactly in an eigenstate, and the dark-state vector undergoes an instantaneous $\pi$ flip without passing through the resonance region. This non-crossing-path design avoids lingering near the minimal energy gap, thereby eliminating the dynamical origin of nonadiabatic transitions.

In contrast, as shown in Fig.~\ref{fig:2}(d), conventional RAP relies on sweeping the detuning $\Delta(t)$ through resonance to drive the system evolution via level crossings. Near $\Delta = 0$, the instantaneous eigenenergy gap $\tilde{\Omega}(t) = \sqrt{\Omega^2(t) + \Delta^2(t)}$ approaches a minimum, where the adiabatic condition is most likely to be violated. Consequently, the system inevitably experiences a finite-duration hazardous region around the resonance point, within which the nonadiabatic coupling remains nonzero. Therefore, a sufficiently large $\Omega_0$ is required to maintain adiabaticity, which inevitably introduces additional errors.

\section{Numerical Results}\label{sec:numerical}

\subsection{Two-qubit Bell state}
\begin{figure*}[htbp]
	\centering
	\includegraphics[width=\textwidth]{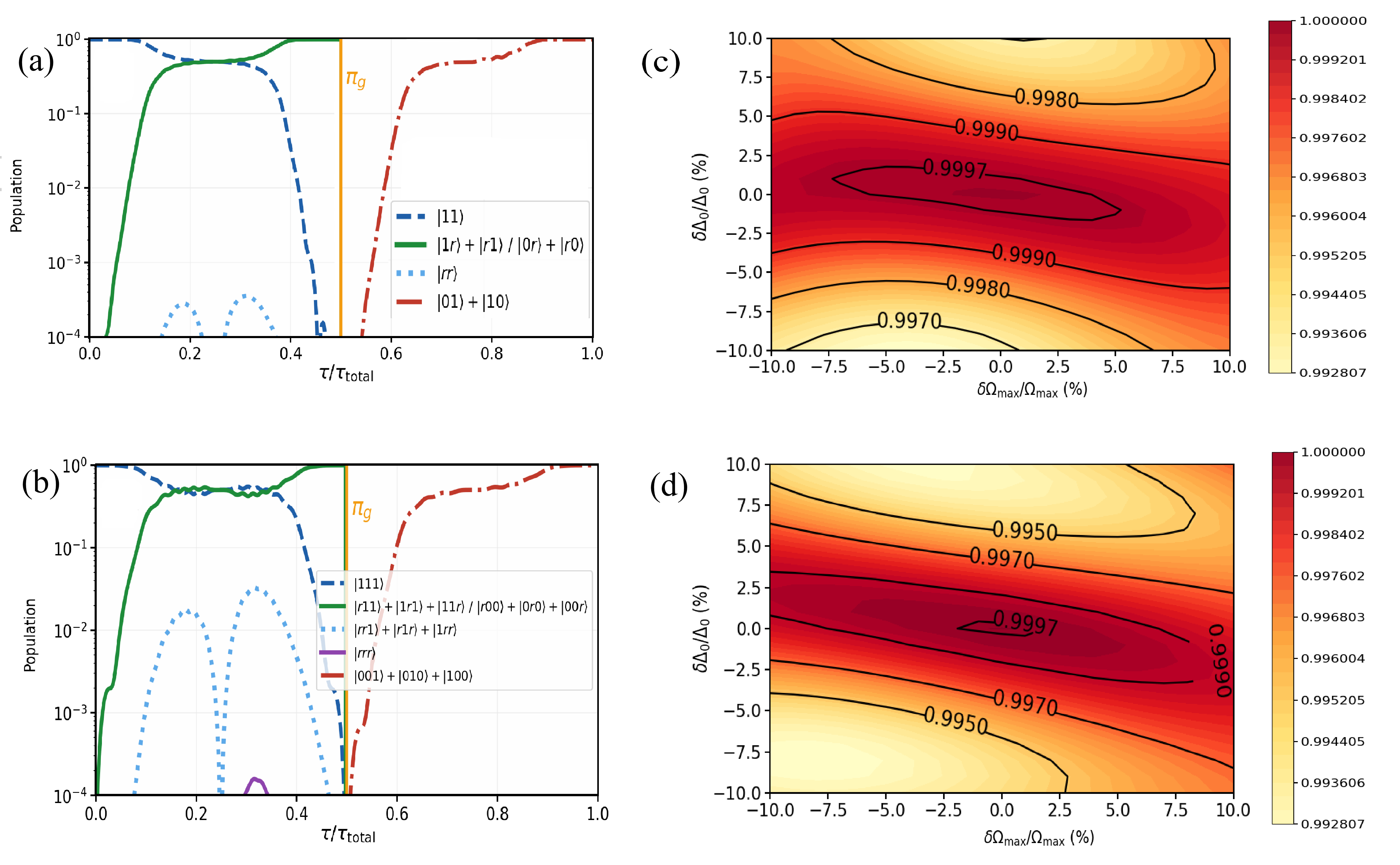}
	\caption{Time evolution and gate fidelity of the two- and three-qubit systems. (a)~Time evolution of the two-qubit system with a blockade strength of $V_0 = 4.12\,\Omega_0$.
		(b)~Time evolution of the three-qubit system with a blockade strength of $V_0 = 4.29\,\Omega_0$.
		(c)~Contour plot of the two-qubit gate fidelity as a function of the control parameters, with $\Omega_0/2\pi = 100\,\mathrm{MHz}$, $\tau_{\text{total}} = 215.7$, $\tau_R = 20.86$, $\tau_D = 78.30$, $\Delta_0 = 1.38\Omega_0$.
		(d)~Contour plot of the three-qubit gate fidelity as a function of the control parameters, with $\Omega_0/2\pi = 100\,\mathrm{MHz}$, $\tau_{\text{total}} = 200.7$, $\tau_R = 20.09$, $\tau_D = 57.81$, $\Delta_0 = 0.865\Omega_0$.}
	\label{fig:3}
\end{figure*}
We first study the preparation of the two-qubit Bell state. The system is initially prepared in the product state $|11\rangle$. The entire preparation process consists of the following four steps. The first level-crossing-free pulse is applied, and under the Rydberg blockade, the system adiabatically evolves to the symmetric single-Rydberg-excitation superposition $(|1r\rangle + |r1\rangle)/\sqrt{2}$. A fast $\pi_g$ pulse is applied to all atoms, achieving population inversion between the ground states $|0\rangle$ and $|1\rangle$ via stimulated Raman transitions or microwave driving \cite{jenkins2022, ma2022, lis2023, levine2019, mcdonnell2022}, while keeping the Rydberg state $|r\rangle$ unchanged. The second level-crossing-free pulse is applied, transferring the single-Rydberg-excitation superposition back to the ground-state entangled superposition, ultimately preparing the target Bell state $(|10\rangle + |01\rangle)/\sqrt{2}$. Numerical simulations employ the full nine-level model, including the ground states $|00\rangle$, $|01\rangle$, $|10\rangle$, $|11\rangle$, the single Rydberg excitation states $|0r\rangle$, $|r0\rangle$, $|1r\rangle$, $|r1\rangle$, and the doubly excited Rydberg state $|rr\rangle$. We identify the optimal parameter combination through systematic optimization of the pulse parameters \cite{jandura2022, dalal2023}: total evolution time $\tau_{\text{total}} = 215.7$, Rabi frequency width $\tau_R = 20.86$, detuning width $\tau_D = 78.30$, peak detuning $\Delta_0 = 1.38\Omega_0$, and blockade strength $V_0 = 4.12\Omega_0$.
\begin{figure*}[htbp]
	\centering
	\includegraphics[width=\textwidth]{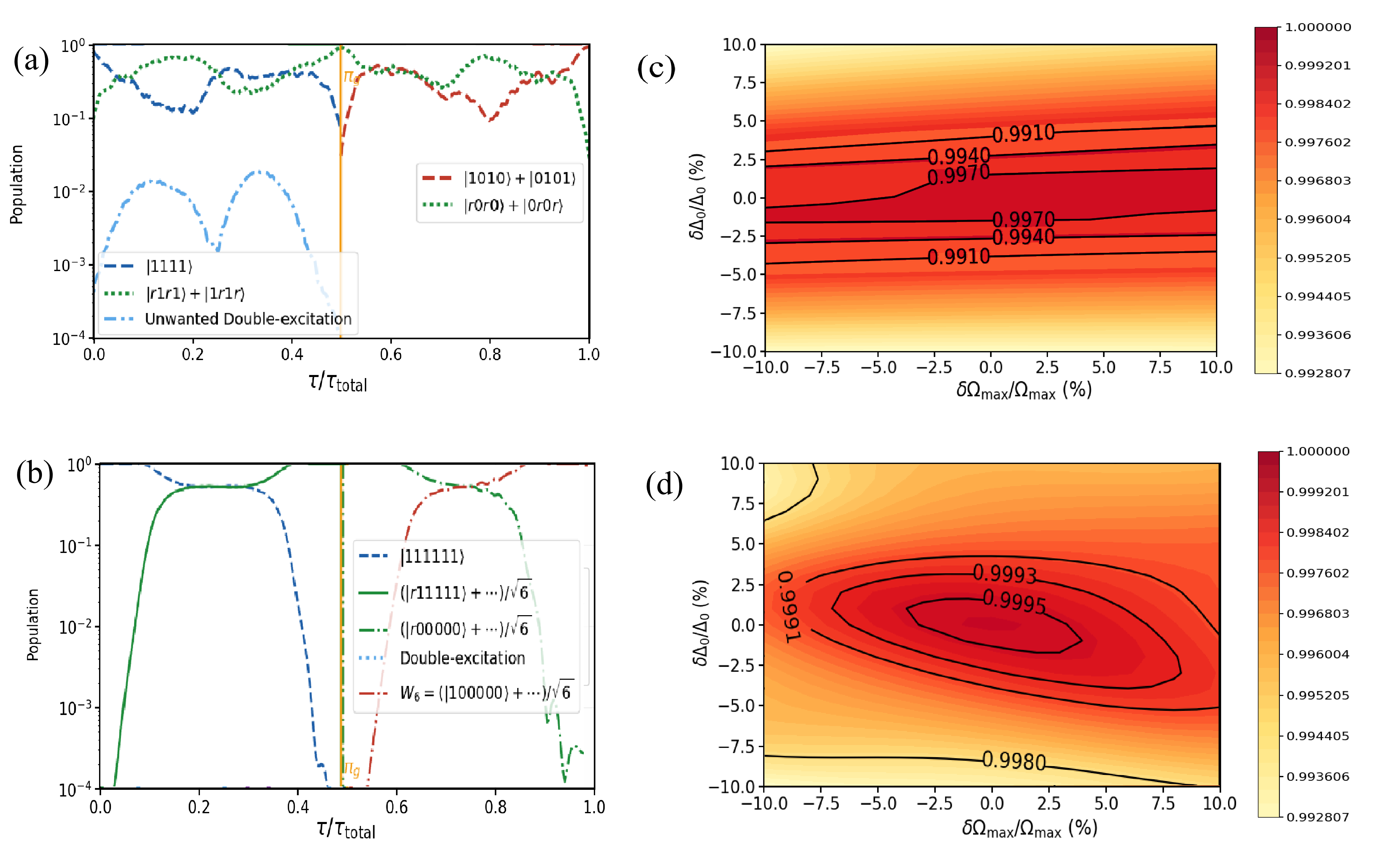}
	\caption{Time evolution and robustness analysis for the four- and six-qubit gates. (a)~Time evolution of the four-qubit system state with a blockade strength of $V_0 = 4.036\Omega_0$.
		(b)~Time evolution of the six-qubit honeycomb system state with a blockade strength of $V_0 = 540\,\mathrm{MHz}$.
		(c)~Robustness of the four-qubit gate fidelity to control-parameter fluctuations, with $\tau_{\text{total}} = 290.0$, $\tau_R = 29.5$, $\tau_D = 40.25$, $\Delta_0 = 0.59\Omega_0$.
		(d)~Robustness of the six-qubit honeycomb gate fidelity to control-parameter fluctuations, with $\tau_{\text{total}} = 221.64$, $\tau_R = 20.80$, $\tau_D = 67.28$, $\Delta_0 = 1.02\Omega_0$.}
	\label{fig:4}
\end{figure*}

We show the population dynamics during the two-qubit Bell state preparation in Fig.~\ref{fig:3}. From the population evolution in Fig.~\ref{fig:3}(a), the system starts from the initial state $|11\rangle$ and gradually transfers to the single Rydberg excitation superposition under the action of the first level-crossing-free pulse. After the $\pi_g$ pulse, the second level-crossing-free pulse transfers the single Rydberg excitation superposition back to the ground-state Bell state, with the final fidelity reaching $F = 0.9997$. Throughout the process, the population of the doubly excited Rydberg state $|rr\rangle$ remains below $10^{-3}$, verifying the effectiveness of strong Rydberg blockade.

Figure.~\ref{fig:3}(c) shows the robustness of the two-qubit Bell state fidelity against pulse parameter fluctuations. Near the optimal parameters ($\delta\Omega_{\max} = \delta\Delta_{0} = 0$), the fidelity reaches $F = 0.9997$. Within $\pm 5\%$ parameter variation, the fidelity remains above $0.9990$; even within $\pm 10\%$ parameter variation, the fidelity remains above $0.997$. The fidelity contours extend along the diagonal direction, indicating a certain correlation between $\Omega_{\max}$ and $\Delta_{0}$, but overall the scheme exhibits excellent robustness.

\subsection{Three-qubit W state}
The three-qubit W-state preparation proceeds similarly to the two-qubit case. The system is initially prepared in the product state $|111\rangle$. The preparation procedure is as follows: the first level-crossing-free pulse drives the system to the three-body single-Rydberg-excitation superposition $(|11r\rangle + |1r1\rangle + |r11\rangle)/\sqrt{3}$ under the Rydberg blockade. A fast $\pi_g$ pulse inverts the ground-state populations. The second level-crossing-free pulse transfers the system to the target W state $(|100\rangle + |010\rangle + |001\rangle)/\sqrt{3}$. Numerical simulations employ the full 27-level model. The optimized pulse parameters are: $\tau_{\text{total}} = 200.7$, $\tau_R = 20.09$, $\tau_D = 57.81$, $\Delta_0 = 0.865\Omega_0$, and blockade strength $V_0 = 4.29\Omega_0$.

We show the population dynamics during the three-qubit W-state preparation in Fig.~\ref{fig:3}(b). Similar to the two-qubit case, the system starts from the initial state $|111\rangle$ and transfers to the three-body single Rydberg excitation superposition $(|11r\rangle + |1r1\rangle + |r11\rangle)/\sqrt{3}$ under the action of the first level-crossing-free pulse. After the $\pi_g$ pulse, the second level-crossing-free pulse transfers the system to the target W state. The final fidelity reaches $F = 0.9997$. The populations of the triply excited Rydberg states remain below $10^{-3}$ throughout the evolution, indicating that the Rydberg blockade is equally effective in the three-qubit system. Compared with the two-qubit system, the three-qubit fidelity is slightly lower, mainly due to numerical error accumulation from the larger Hilbert space and more complex energy level structures.

Figure.~\ref{fig:3}(d) shows the robustness of the three-qubit W state fidelity against parameter fluctuations. Near the optimal parameters, the fidelity reaches $F = 0.9997$. The fidelity contours exhibit an elliptical distribution, with the long axis along the $\delta\Omega_{\max}$ direction and the short axis along the $\delta\Delta_{0}$ direction, indicating that the scheme is more sensitive to fluctuations in $\Delta_{0}$. Within $\pm 3\%$ parameter variation, the fidelity remains above $0.997$; within $\pm 10\%$ variation, the fidelity remains above $0.995$. This demonstrates that the zero-area pulse scheme also possesses excellent robustness in the three-qubit system.

In the three-qubit W system, the advantage of the zero-area pulse is even more pronounced, because the three-body energy level structure is more complex, with more potential level crossing points, and the nonadiabatic coupling effects in conventional RAP schemes are more significant \cite{khazali2020, bosch2023, shao2023}. The zero-area pulse effectively eliminates these additional loss channels by avoiding all level crossings.

\subsection{Four-qubit GHZ state}

We subsequently investigated the preparation of the four-qubit GHZ state. For the four-qubit GHZ state, we adopt the spatial correlation strategy. Four atoms are arranged in a square configuration, with blockade strength $V_0$ between adjacent atoms and interaction $V_1 = V_0/(\sqrt{2})^6 = V_0/8$ between diagonal atoms\cite{walker2008, zwierz2009}. The preparation process consists of the following steps, starting from $|1111\rangle$. The first level-crossing-free pulse prepares the intermediate state $|1r1r\rangle + |r1r1\rangle$ through the selective excitation enabled by the strong blockade between adjacent atoms and the weak interaction between diagonal atoms. A fast $\pi_g$ pulse is applied. The second level-crossing-free pulse transfers the intermediate state to the target GHZ state $(|0101\rangle + |1010\rangle)/\sqrt{2}$. This strategy exploits the significant difference in interaction strength between adjacent and diagonal atoms, selectively transferring the population to the target state \cite{xu2024, gujarati2018, graham2022, bluvstein2022, radnaev2025}. Physically, the strong blockade between adjacent atoms prevents simultaneous excitation of two adjacent atoms to the Rydberg state; however, the weak interaction between diagonal atoms allows diagonal atoms to be simultaneously excited. Therefore, the system evolves to the specific doubly excited state $|1r1r\rangle + |r1r1\rangle$, rather than all possible doubly excited superpositions. This selective evolution is the core of the spatial correlation strategy and the key to achieving high-fidelity GHZ state preparation. Numerical simulations employ the full 81-level model. The target state is the GHZ state $(|0101\rangle + |1010\rangle)/\sqrt{2}$.

Figure.~\ref{fig:4}(a) shows the preparation process of the four-qubit GHZ state. Unlike the Bell and W states, the preparation of the GHZ state requires exploiting the spatial correlations between atoms. In the first step, the system starts from $|1111\rangle$ and prepares the intermediate state $|1r1r\rangle + |r1r1\rangle$ through the first level-crossing-free pulse and the $\pi_g$ pulse. The selective preparation of this intermediate state relies on the strong blockade between adjacent atoms $V_0$ and the weak interaction between diagonal atoms $V_1$: adjacent atoms cannot be simultaneously excited to the Rydberg state, while diagonal atoms can. In the second step, the second level-crossing-free pulse transfers the intermediate state to the target GHZ state. The population of non-target doubly excited states is effectively suppressed during the evolution, verifying the effectiveness of the spatial correlation strategy.

The robustness of the four-qubit GHZ state fidelity against pulse parameter fluctuations is shown in Fig .~\ref{fig:4}(c). At the optimal parameters, the maximum fidelity reaches $F_{\max} = 0.997$. Within $\pm 5\%$ parameter variation, the fidelity remains above $0.99$, and in the central region the fidelity exceeds $0.995$. Compared with the two-qubit and three-qubit systems, the fidelity contour distribution of the four-qubit system is more uniform, possibly because the larger parameter space smooths the fidelity variation. Although the absolute fidelity of the four-qubit system is slightly lower than that of the two-qubit and three-qubit systems, this is mainly due to numerical integration errors in the 81-dimensional Hilbert space and the more complex interaction network. Overall, the scheme still exhibits good performance and robustness in the four-qubit system.

Compared with conventional RAP-based protocols for generating four-qubit GHZ states, our scheme achieves a significantly higher fidelity. This is mainly because the four-qubit system has a larger Hilbert space and a more complex energy level structure, and multiple level crossings in conventional RAP schemes interfere with each other, leading to population leakage to non-target states \cite{li2017, li2019}. The zero-area pulse avoids this problem by not involving any level crossings.

\subsection{Six-qubit honeycomb W state}

Finally, we investigated the preparation of the six-qubit honeycomb W state \cite{li2020,kobayashi2024,evered2025,gypens2018,peregbarnea2012}. The system is prepared in the product state $|111111\rangle$. The protocol mirrors that for fewer qubits, differing only in the atomic spatial configuration and interaction strengths. Six atoms are arranged in a regular hexagon, with nearest-neighbor blockade strength $V_0$, next-nearest-neighbor interaction $V_1 = V_0/27$, and opposite-atom interaction $V_2 = V_0/64$. The preparation process consists of the following steps: the first level-crossing-free pulse drives the system, under the Rydberg blockade, into the single-excitation Rydberg superposition $(|r11111\rangle + |1r1111\rangle + \dots + |11111r\rangle)/\sqrt{6}$. A fast $\pi_g$ pulse flips the ground-state populations. The second level-crossing-free pulse maps this superposition onto the target six-qubit honeycomb W state $(|100000\rangle + |010000\rangle + \dots + |000001\rangle)/\sqrt{6}$. The optimized pulse parameters are: $\tau_{\text{total}} = 221.64$, $\tau_R = 20.80$, $\tau_D = 67.28$, $\Delta_0 = 1.02\Omega_0$.

We show the population dynamics during the six-qubit honeycomb W-state preparation in  Fig.~\ref{fig:4}(b). The system evolves from the initial state $|111111\rangle$ to the six-body single-Rydberg superposition under the first non-level-crossing pulse. Following the $\pi_g$ pulse at $\tau = \tau_{\text{total}}/2$, the second pulse transfers the population to the target W state. The double-excitation population stays below $10^{-3}$ throughout, demonstrating robust Rydberg blockade and the selectivity afforded by the hierarchical interaction strengths. Figure.~\ref{fig:4}(d) shows the robustness of the six-qubit honeycomb W-state fidelity to pulse parameter fluctuations. At the optimal point, the peak fidelity reaches $F_{\max} = 0.9995$. The fidelity remains above $0.999$ within $\pm5\%$ parameter variations and exceeds $0.995$ even for $\pm10\%$ fluctuations.

The preparation fidelity of the six-qubit honeycomb W state remains higher than that of the four-qubit GHZ state, despite the increase in system size. This improvement stems from the fact that W-state preparation avoids the complicated spatial selection required for GHZ states, while fully exploiting the level-crossing-free nature of the pulse. In larger systems, conventional RAP suffers severely from level-crossing-induced diabatic errors; our scheme, by remaining adiabatically robust throughout the protocol, becomes comparatively even more advantageous. Although the fidelity of the six‑qubit W state is slightly lower than that of the three‑qubit W state, the degradation is negligible, with the fidelity remaining above $0.9995$. The experimental realization of the six‑qubit honeycomb W state demonstrates the scalability of our level‑crossing‑free RAP scheme to extended atomic arrays, which is crucial for large‑scale systems characterized by dense and complex spectra.

\section{Generalization And Conclusion}\label{sec:generalization and conclusion}

In summary, we have proposed a rapid adiabatic passage scheme based on level-crossing-free pulses for efficiently generating multiqubit entangled states in Rydberg atom systems. The scheme employs antisymmetric Rabi frequency pulses and even-symmetric detuning pulses, enabling the system to completely avoid level crossings throughout the evolution and thereby fundamentally eliminating the risk of nonadiabatic transitions near level crossing points. By exploiting the Rydberg blockade effect, two sequential level-crossing-free pulses interspersed with a $\pi_g$ pulse prepare two-qubit Bell states, three-qubit W states, four-qubit GHZ states, and six-qubit honeycomb W states from an initial product state with high fidelity.

Numerical simulations show that the fidelities exceed $0.9997$ for the two-qubit Bell state, $0.9997$ for the three-qubit W state, $0.997$ for the four-qubit GHZ state, and $0.9995$ for the six-qubit honeycomb W state. 
Although the fidelity of the six-qubit honeycomb W state is slightly lower than that of its three-qubit W state, this degradation is negligible, remaining well above $0.9995$. The preparation of this six-qubit honeycomb W state provides compelling evidence for the scalability of our level-crossing-free RAP protocol in extended atomic arrays---a critical advantage when dealing with large-scale systems characterized by dense and complex spectra. 
The superior fidelity of the six-qubit honeycomb W state over its four-qubit GHZ counterpart arises not only from the elimination of intricate spatial-selection constraints but also from a much simpler effective dynamics despite the larger Hilbert space. 
In contrast to conventional RAP, where multiple level crossings induce severe diabatic errors in many-body regimes, our level-crossing-free protocol remains inherently robust, with its relative performance gain becoming even more pronounced as the system scale grows.

In our scheme, $\Delta(t)$ maintains a single sign and only instantaneously touches the resonance point at $t=0$. This is enabled by the symmetric design of the level-crossing-free pulse, under which the system resides strictly in an eigenstate at that instant, thereby eliminating the need for the system to undergo the slow adiabatic passage through the resonance region as required in the conventional RAP. This mechanism fundamentally removes the dynamical cause of non-adiabatic transitions, leading to highly efficient and robust adiabatic evolution. This scheme provides a simple, efficient, and robust method for entangled state preparation in neutral-atom quantum information processing, with potential for extension to larger-scale atomic arrays, and is expected to be applied in quantum computing, quantum simulation, and quantum networks \cite{semeghini2021, ebadi2022, evered2025, wilson2022, burgers2022, cao2024}.

\section*{Acknowledgments}\label{sec:Acknowledgments}
This work was supported by the National Natural Science Foundation of China (Nos. 12574375 and 12004231), the Fundamental Research Program of Shanxi Province (Grant No. 202203021211301), and the Research Project Supported by Shanxi Scholarship Council of China (Nos. 2023‑028 and 2022‑014).

\bibliographystyle{apsrev4-2}
\bibliography{refs}

\end{document}